\documentclass[conference]{IEEEtran}
\usepackage{graphicx}
\usepackage{amsmath}

\ifCLASSINFOpdf
\else
\fi
\begin{document}
%
\title{Impact of credit default swaps on financial contagion}

\author{
\IEEEauthorblockN{Yoshiharu Maeno\\ and Satoshi Morinaga}
\IEEEauthorblockA{NEC Corporation\\
Kawasaki-shi, Kanagawa 211-8666, Japan\\
Email: y-maeno@aj.jp.nec.com\\
Email: morinaga@cw.jp.nec.com}
\and
\IEEEauthorblockN{Kenji Nishiguchi}
\IEEEauthorblockA{Japan Research Institute\\
Shinagawa-ku, Tokyo 141-0022, Japan\\
Email: nishiguchi.kenji@jri.co.jp}
\and
\IEEEauthorblockN{Hirokazu Matsushima}
\IEEEauthorblockA{Institute for International\\Socio-economic Studies\\
Minato-ku, Tokyo 108-0073, Japan\\
Email: h-matsushima@ah.jp.nec.com}
}

%


\maketitle

\begin{abstract}
It had been believed in the conventional practice that the risk of a bank going bankrupt is lessened in a straightforward manner by transferring the risk of loan defaults. But the failure of American International Group in 2008 posed a more complex aspect of financial contagion. This study presents an extension of the asset network systemic risk model (ANWSER) to investigate whether credit default swaps mitigate or intensify the severity of financial contagion. A protection buyer bank transfers the risk of every possible debtor bank default to protection seller banks. The empirical distribution of the number of bank bankruptcies is obtained with the extended model. Systemic capital buffer ratio is calculated from the distribution. The ratio quantifies the effective loss absorbency capability of the entire financial system to force back financial contagion. The key finding is that the leverage ratio is a good estimate of a systemic capital buffer ratio as the backstop of a financial system. The risk transfer from small and medium banks to big banks in an interbank network does not mitigate the severity of financial contagion.
\end{abstract}


%
\IEEEpeerreviewmaketitle

\section{Introduction}

Understanding how the characteristics of a financial system govern the financial contagion of bank bankruptcies is essential in the argument to reform the capital requirement and other regulatory standards. Recently computer simulation models are developed to mimic the transmission of financial distress and predict the severity of financial contagion\cite{Roukny}, \cite{Arinaminpathy}, \cite{Gai11}, \cite{Haldane}, \cite{Upper}, \cite{Gai}, \cite{Jaramillo}, \cite{May}, \cite{Gatti}. Both the external assets and interbank loans of banks can be the origin of financial distress in these models. Either distress may transmit separately in a peace time while compound distress transmits in a crisis time. A bank makes an investment in multiple external asset classes. The value of the total external assets may go downturn when the markets fluctuate. A defective investment portfolio of banks imposes financial distress on them\cite{Beale}. A failing debtor bank becomes insolvent in paying off the interbank borrowings. Any creditor banks suffer financial distress from the failing debtor bank\cite{Nier}. A bank goes bankrupt unless the capital buffer absorbs the total loss from the external assets and interbank loans. Bank bankruptcies bring about still more financial distress repeatedly. This is the mechanism of financial contagion.

It had been believed in the conventional practice that the risk of a bank going bankrupt is lessened in a straightforward manner by transferring the risk of loan defaults. But the failure of American International Group in 2008 posed a more complex aspect of financial contagion. AIG had sold protection in the form of credit default swaps to insure securities worth \$441 billion. The value of the securities declined as the subprime loan market collapsed. The credit rating of AIG was downgraded. AIG faced a deadly liquidity crisis. The bankruptcy of AIG would have caused catastrophic damage to the financial system. The Fed proposed a rescue package and the crisis ended in the unprecedented taxpayer-financed bailout of a giant private company.

The market for CDS has been growing\cite{Battiston} despite the bankers' painful awareness that CDS are imperfect. CDS are financial instruments for risk transfer\cite{Namatame} which relate three banks. A protection seller bank compensates the loss a protection buyer bank incurs in the specified credit event of a third-party reference bank. Risk transfer is meant to level off the risk of individual banks. But once a severe financial distress strikes a key seller like AIG, risk transfer does not work any longer. A fragile financial system appears abruptly. Systemic risk\cite{Helbing} is rather intensified.

This study presents an extension of the asset network systemic risk model (ANWSER)\cite{Maeno13a}, \cite{Maeno} to investigate whether the CDS mitigate or intensify the severity of financial contagion. A protection buyer bank transfers the risk of every possible debtor bank default to protection seller banks. The protection buyer bank may make additional loans and lend as much as the risk-transferred interbank loans. The empirical distribution of the number of bank bankruptcies is obtained under these conditions with the extended model. Systemic capital buffer ratio is calculated from the distribution. The ratio quantifies the effective loss absorbency capability of the entire financial system to force back financial contagion. The extended model demonstrates how the systemic capital buffer ratio is affected by the denseness and concentration of the interbank network and related to the core tier 1 ratio and the leverage ratio.

\section{Model}
\label{Model}

\begin{figure}
\centering
\includegraphics[width=1.8in, angle=-90]{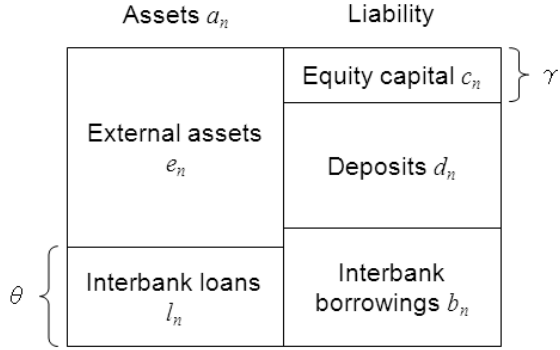}
\caption{Balance sheet of the $n$-th bank. The balance sheet consists of interbank loans $l_{n}$, external assets $e_{n}$, equity capital $c_{n}$, interbank borrowings $b_{n}$, and deposits $d_{n}$.}
\label{201309301}
\end{figure}

\begin{figure}
\centering
\includegraphics[width=2.2in, angle=-90]{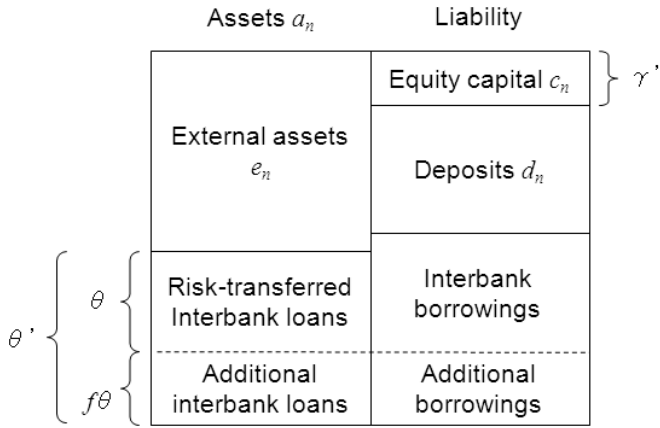}
\caption{Balance sheet of the $n$-th bank after the risk transfer of interbank loans and additional loans. The value of the additional interbank loans as a fraction of the value of the risk-transferred interbank loans is $f$.}
\label{201309302}
\end{figure}

\begin{figure}
\centering
\includegraphics[width=3in, angle=-90]{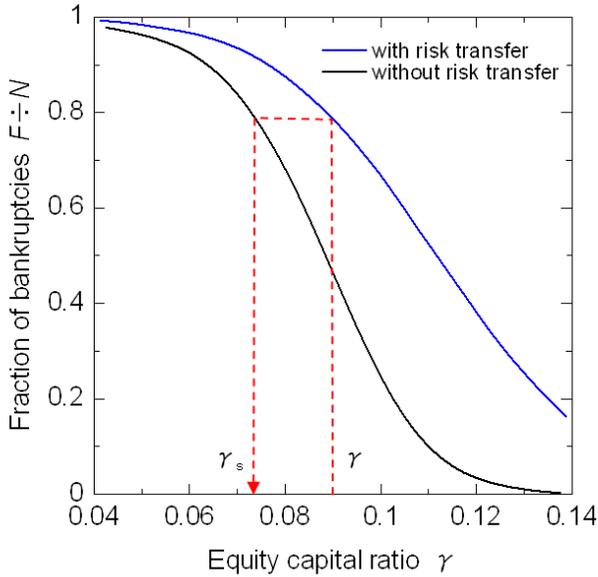}
\caption{Systemic capital buffer ratio $\gamma_{s}$ which is calculated as a function of $\gamma$ by comparing the number of bank bankrupcies in financial systems with risk transfer and without risk transfer.}
\label{201309309}
\end{figure}

Models of interbank loans, investments, and CDS are presented in this section.

The asset network systemic risk model (ANWSER)\cite{Maeno13b} is founded on previous computer simulation models\cite{Beale}, \cite{Nier}. They investigate the statistical characteristics of a financial system with a Monte-Carlo method. The Monte-Carlo method is a broad class of a computational technique to obtain many samples of numerical outcomes which are used to analyze the statistical characteristics. The technique relies on a sequence of random numbers generated repeatedly from a specified probability distribution. The initial financial distress on banks is the falling prices of their external assets in the market. When a debtor bank happens to go bankrupt, the consequent interbank loan defaults are the next financial distress to its creditor banks. Financial distress transmits from failing debtor banks to creditor banks repeatedly in an interbank network.

$N$ is the number of banks. $M$ is the number of external assets in which an individual bank makes an investment. Fig. \ref{201309301} shows the balance sheet of the $n$-th bank. The interbank loan ratio $\theta = \sum_{n=1}^{N} l_{n} / \sum_{n=1}^{N} a_{n}$ is the total value of interbank loans as a fraction of the total value of assets. The assets consist of the interbank loans $l_{n}$ and external assets $e_{n}$ like securities and government bonds. An interbank loan is the credit relation between a creditor bank and a debtor bank which appears when the debtor bank raises money in the interbank market. A interbank network describes the all credit relations. It is a directed graph which consists of banks as vertices, and the interbank loans as edges from creditor banks to debtor banks.

Both big banks and small banks have the same equity capital ratio $\gamma = c_{n} / a_{n}$. It is the same as the core tire 1 ratio $t=\gamma$ and the leverage ratio $l=\gamma$. This value is the minimal level of the equity capital ratio required by the bank regulatory policies. The liability consists of the equity capital $c_{n}$, interbank borrowings $b_{n}$, and deposits $d_{n}$. The equity capital includes common stock and disclosed reserves. These need not be paid off and can be used to absorb the loss from financial distress immediately.

The denseness $\kappa$ of the financial system is the average incoming or outgoing nodal degree of the interbank network as a fraction of $N-1$. A more dense interbank credit network has a larger value of $\kappa$. The concentration $\rho$ of the financial system is the sum of the interbank loan share of the 1 percent top banks (the largest, the second largest, $\cdots$, the $N/100$-th largest banks). A more concentrated interbank credit network has a larger value of $\rho$.

Every creditor bank buys protection in the form of CDS which refer to the all of its debtor banks. Every interbank loan default is compensated by one of protection seller banks. $S \leq N$ is the number of protection seller banks. A protection seller bank can be a creditor and debtor bank in the interbank network. The risk weight of these interbank loans becomes 0\%. After this risk transfer, every creditor bank makes a borrowing and lends additionally as much as the value of the interbank loans whose risk was transferred. The risk of the additional interbank loans is not transferred. Their risk weight is 100\%. The parameter $f \geq 0$ specifies the value of the additional interbank loans as a fraction of the value of the interbank loans whose risk was transferred. Fig. \ref{201309302} shows the balance sheet after the risk transfer and additional loans.

The core tire 1 ratio $t'$ is given by eq.(\ref{t'}). The core tier 1 ratio is calculated based on the risk weighted assets.
\begin{equation}
t'= \frac{\gamma}{1-\theta+f\theta}.
\label{t'}
\end{equation}

The leverage ratio $l'$ is given by (\ref{l'}). The leverage ratio is calculated based on the value of the equity capital against the overall assets regardless of the risk. 
\begin{equation}
l'= \frac{\gamma}{1+f\theta}.
\label{l'}
\end{equation}

Given $N$, $M$, and $S$, a sequence of random numbers is generated to synthesize samples for fixed values of $\theta$, $\gamma$, $\kappa$, $\rho$, and $f$. An individual sample includes
\begin{itemize}
\item interbank network topology $\mbox{\boldmath{$Z$}}$ (an $N \times N$ matrix) where the element $Z_{nn'}=1$ means the $n'$-th bank makes a loan from the $n$-th bank and otherwise $Z_{nn'}=0$ 
\item risk transfer pattern $\mbox{\boldmath{$Y$}}$ (an $N \times N \times N$ tensor) where the element $Y_{nn'n''}=1$ means the $n$-th bank buys protection for the loan default of the $n'$-th bank from the $n''$-th bank and otherwise $Y_{nn'n''}=0$
\item investment portfolio $\mbox{\boldmath{$X$}}$ (an $N \times M$ matrix) where the element $X_{nm}$ is the fraction of the investment which the $n$-th bank makes in the $m$-th external asset class ($\sum_{m=1}^{M} X_{nm} =1$, $0 \leq X_{nm} \leq 1$)
\item prices of the external assets in the market $\mbox{\boldmath{$v$}}$ (an $M$ column vector) where the element $v_{m}$ is the price of the unit of the $m$-th external asset class.
\end{itemize}

The initial financial distress on the $n$-th bank is $e_{n} \sum_{m=1}^{M} X_{nm} v_{m}$. 

A protection buyer bank goes bankrupt if the total loss from the financial distress is not absorbed by its capital buffer. As far as the protection buyer bank survives, the loss in capital buffer in the event of debtor bank bankruptcies is compensated by the CDS payoff from the protection seller banks. The protection seller bank goes bankrupt if the sum of the total loss from the financial distress and the CDS payoff is not absorbed by its capital buffer. It is assumed that failing debtor banks do not pay off any portions of the interbank loans to creditor banks. $F$ is the number of banks which end in bankruptcy until the financial contagion comes to a halt. The empirical distribution of the number of bank bankruptcies $P(F)$ is obtained from those samples. The value of $F$ is picked up at the 999-th 1000-quantile point as the representative in case of a financial crisis.

Systemic capital buffer ratio $\gamma_{s}$ is calculated by comparing the curves for the number of bank bankruptcies $F$ as a function of $\gamma$ in financial systems with risk transfer and without risk transfer. The systemic capital buffer ratio refers to the equity capital ratio in the financial system without risk transfer at which the number of bank bankruptcies is the same as that in the financial system with risk transfer. Fig. \ref{201309309} shows how to calculate $\gamma_{s}$ as a function of $\gamma$. In this case, $\gamma_{{\rm s}}=0.073$ when $\gamma=0.09$. A negative impact is meant if $\gamma_{{\rm s}} < \gamma$.

\section{Experimental condition}
\label{Experimental}

\begin{figure}
\centering
\includegraphics[width=2.3in, angle=-90]{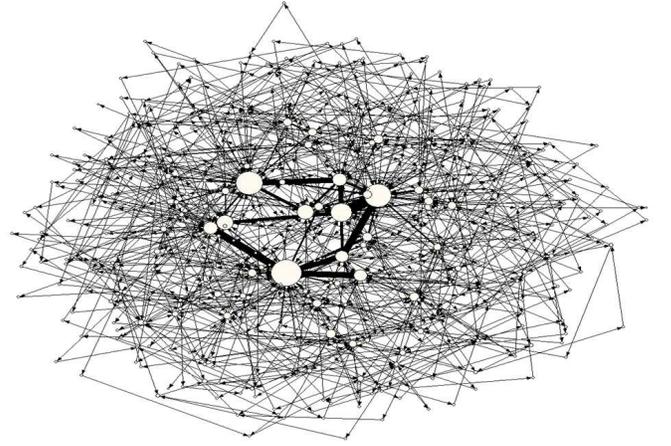}
\caption{Interbank network of $N=500$. The size of a vertex represents the value of assets of a bank. The width of an edge represents the value of an interbank loan between the banks at its ends.}
\vspace{0.2in}
\label{1124b4}
\end{figure}

The experimental conditions are presented in this section.

Two cases are compared to investigate the impact of the CDS with the model in \ref{Model}. In case that the risk transfer is absent, the number of bank bankruptcies is measured for the given value of the interbank loan ratio $\theta$ under the condition $\mbox{\boldmath{$Y$}}=\mbox{\boldmath{$0$}}$. In case that the risk transfer is present, the number is measured for the calculated value of the interbank loan ratio $\theta'=(\theta + f\theta)/(1+f\theta)$ in Fig.\ref{201309302}. In both cases, the values of four parameters are fixed. $N=500$, $M=2$, $S=10$, and $\theta=0.3$. The values of the remaining parameters can be adjusted in the range of $0.04 \leq \gamma \leq 0.14$, $0.01 \leq \kappa \leq 0.1$, $0.1 \leq \rho \leq 0.5$, and $0 \leq f \leq 1$. The probability distributions for random variables $\mbox{\boldmath{$Z$}}$, $\mbox{\boldmath{$Y$}}$, $\mbox{\boldmath{$X$}}$, and $\mbox{\boldmath{$v$}}$ are assumed as follows in this study. 

Funds transfer between banks are found highly heterogeneous in CHAPS of the United Kingdom\cite{Becher}, e-MID of Italy\cite{Masi}, Fedwire of the United States\cite{Soramaki}, BOJ-Net of Japan\cite{Inaoka}, and worldwide\cite{Minoiu}. The nodal degree of the network and the value of the transferred funds obey a power law. In this study, $\mbox{\boldmath{$Z$}}$ is generated randomly by a generalized Barab\'{a}si-Albert model\cite{Barabasi}, \cite{Dorogovtsev}. This is a random graph with the mechanism of growth and preferential attachment which becomes scale-free as $N$ goes to infinity. The distribution of the nodal degree $k$ obeys the power law $P(k) \propto k^{-\alpha}$ where $\alpha \geq 2$. There is a significant probability of the presence of very big banks. This is the origin of heterogeneity. Fig. \ref{1124b4} shows an example of an interbank network of $N=500$.

The value of a loan from the $n$-the bank to the $n'$-th bank is determined from the incoming nodal degree $k^{{\rm in}}_{n} = \sum_{n'=1}^{N} Z_{n'n}$ and outgoing nodal degree $k^{{\rm out}}_{n} = \sum_{n'=1}^{N} Z_{nn'}$ of the interbank network topology by the generalized law in eq.(\ref{wnn'}). The concentration $\rho$ increases as $r \geq 0$ increases. The value of interbank loans is a constant if $r=0$.
\begin{equation}
w_{nn'} \propto \frac{ Z_{nn'} \, (k_{n}^{{\rm out}} \, k^{{\rm in}}_{n'})^r }{\sum_{n \neq n'} Z_{nn'} \, (k_{n}^{{\rm out}} \, k^{{\rm in}}_{n'})^r }.
\label{wnn'}
\end{equation}

Once the value of $w_{nn'}$ is determined, the interbank loans and borrowings are given by eq.(\ref{ln}) and (\ref{bn'}). 
\begin{equation}
l_{n} = \sum_{n'} w_{nn'}.
\label{ln}
\end{equation}
\begin{equation}
b_{n'} = \sum_{n} w_{nn'}.
\label{bn'}
\end{equation}

The external assets are given by eq.(\ref{en}). A prerequisite that the external assets are no less than the net interbank borrowings ($e_{n}+l_{n} \geq b_{n}$) is imposed because the bank has already gone bankrupt if this prerequisite is not satisfied. Then $c_{n}=\gamma(e_{n}+l_{n})$ and $d_{n}=e_{n}+l_{n}-c_{n}-b_{n}$.
\begin{eqnarray}
e_{n} &=& \max(b_{n}-l_{n}, 0) \nonumber \\
&+& (\frac{1-\theta}{\theta} \sum_{n=1}^{N} l_{n} - \sum_{n=1}^{N} \max(b_{n}-l_{n}, 0)) \frac{l_{n}}{\sum_{n=1}^{N} l_{n}}. \nonumber \\
\label{en}
\end{eqnarray}

Every creditor bank buys protection for its debtor bank defaults. Protection seller banks are the largest $S=10$ banks in the value of assets. The risk is transferred from small and medium banks to big banks. A protection buyer bank chooses one protection seller bank for individual loans randomly.

A bank chooses multiple external asset classes to make an investment in randomly. When $M=2$, $X_{n1}$ and $X_{n2}$ obey a uniform distribution. The prices of the external asset classes are independently and identically distributed. The absolute fluctuation in their prices obey a uni-variate Student $t$-distribution. The prices rise or fall randomly. The degree of freedom is $\mu=1.5$. This is a long tailed distribution which is suitable to describe a sudden large fluctuation. The amplitude of the absolute fluctuation is adjusted so that the probability of a bank with the equity capital ratio $\gamma=0.07$ alone going bankrupt can be $p=10^{-3}$.  

\section{result}

\begin{figure}
\centering
\includegraphics[width=3in, angle=-90]{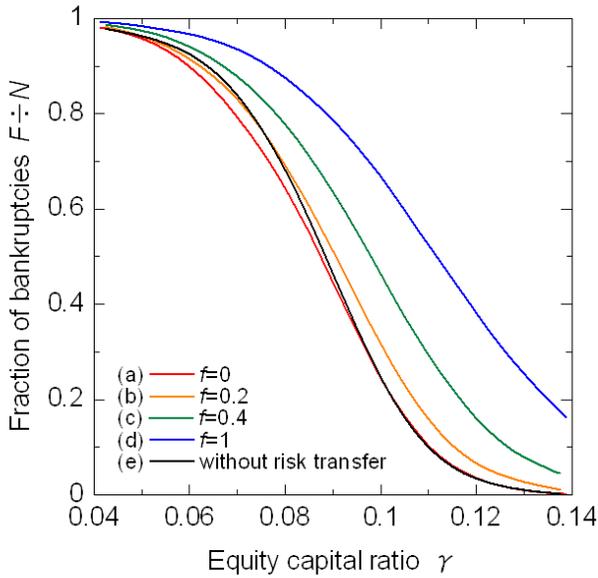}
\caption{Number of bank bankruptcies as a fraction of $N$ as a function of $\gamma$ for $\kappa=0.05$, $\rho=0.3$ (a baseline interbank network), and $f=0$, 0.2, 0.4, 1 when $N=500$, $M=2$, $S=10$, and $\theta=0.3$.}
\label{201309303}
\end{figure}

\begin{figure}
\centering
\includegraphics[width=3in, angle=-90]{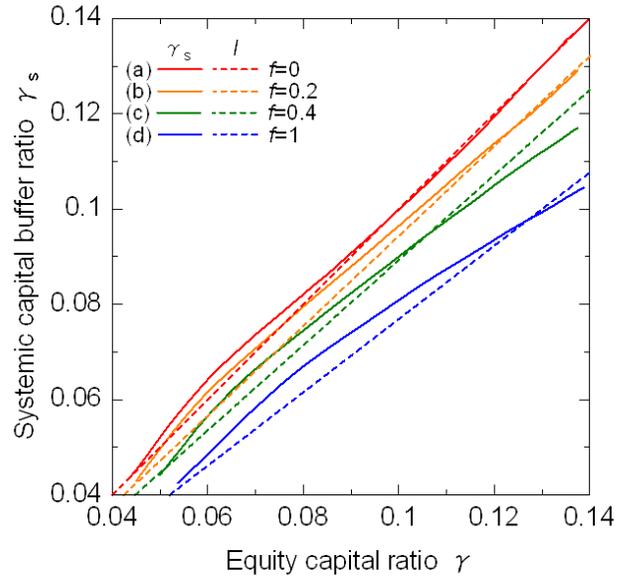}
\caption{Systemic capital buffer ratio $\gamma_{s}$ as a function of $\gamma$ in solid lines for $\kappa=0.05$, $\rho=0.3$, and $f=0$, 0.2, 0.4, and 1 when $N=500$, $M=2$, $S=10$, and $\theta=0.3$. The leverage ratio $l'$ for $f=0$, 0.2, 0.4, and 1 are shown by broken lines.}
\vspace{0.2in}
\label{201309304}
\end{figure}

\begin{figure}
\centering
\includegraphics[width=3in, angle=-90]{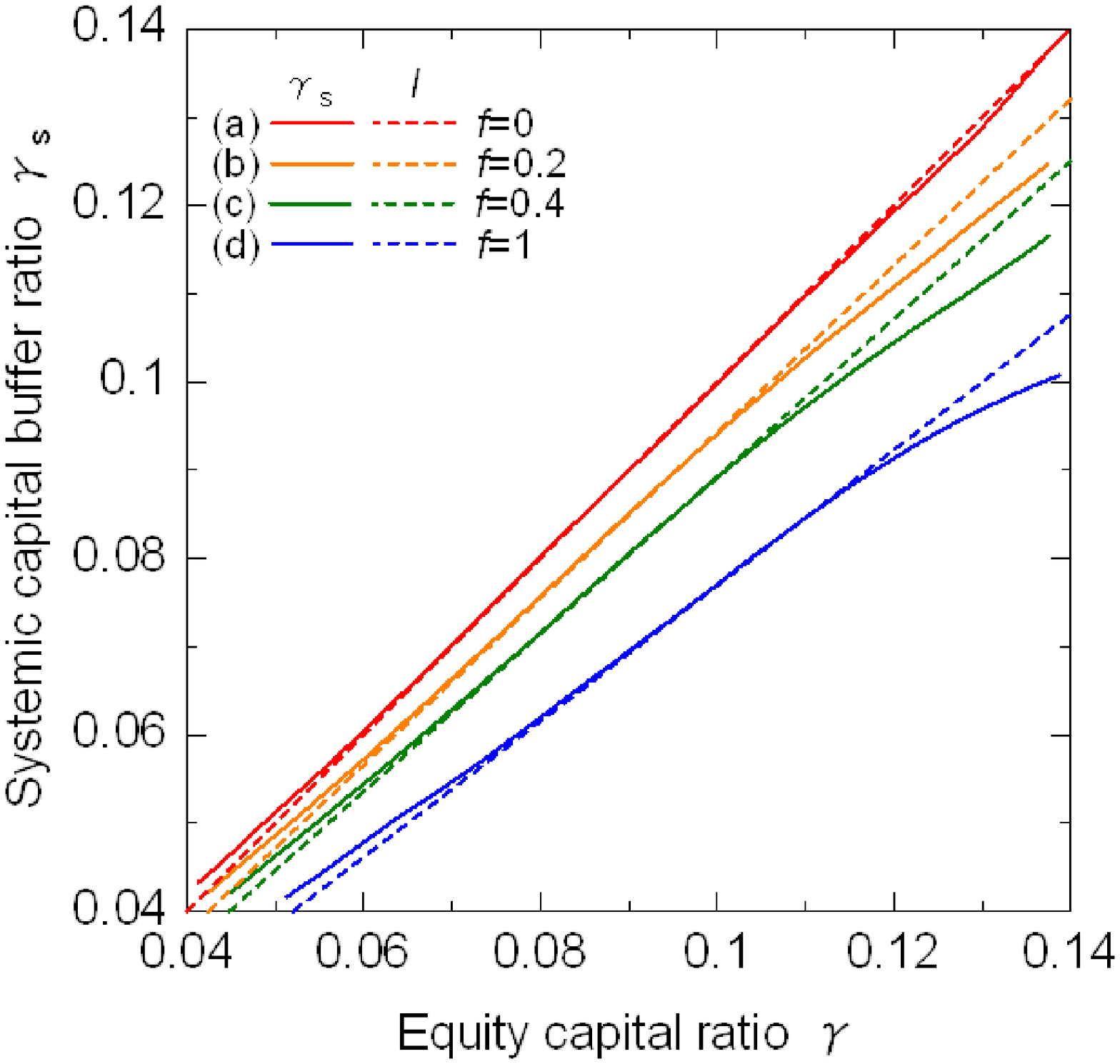}
\caption{Systemic capital buffer ratio $\gamma_{s}$ as a function of $\gamma$ for $\kappa=0.01$, $\rho=0.3$ (a less strongly connected interbank network), and $f=0$, 0.2, 0.4, 1 when $N=500$, $M=2$, $S=10$, and $\theta=0.3$. The leverage ratio $l'$ are shown by broken lines.}
\label{201309305}
\end{figure}

\begin{figure}
\centering
\includegraphics[width=3in, angle=-90]{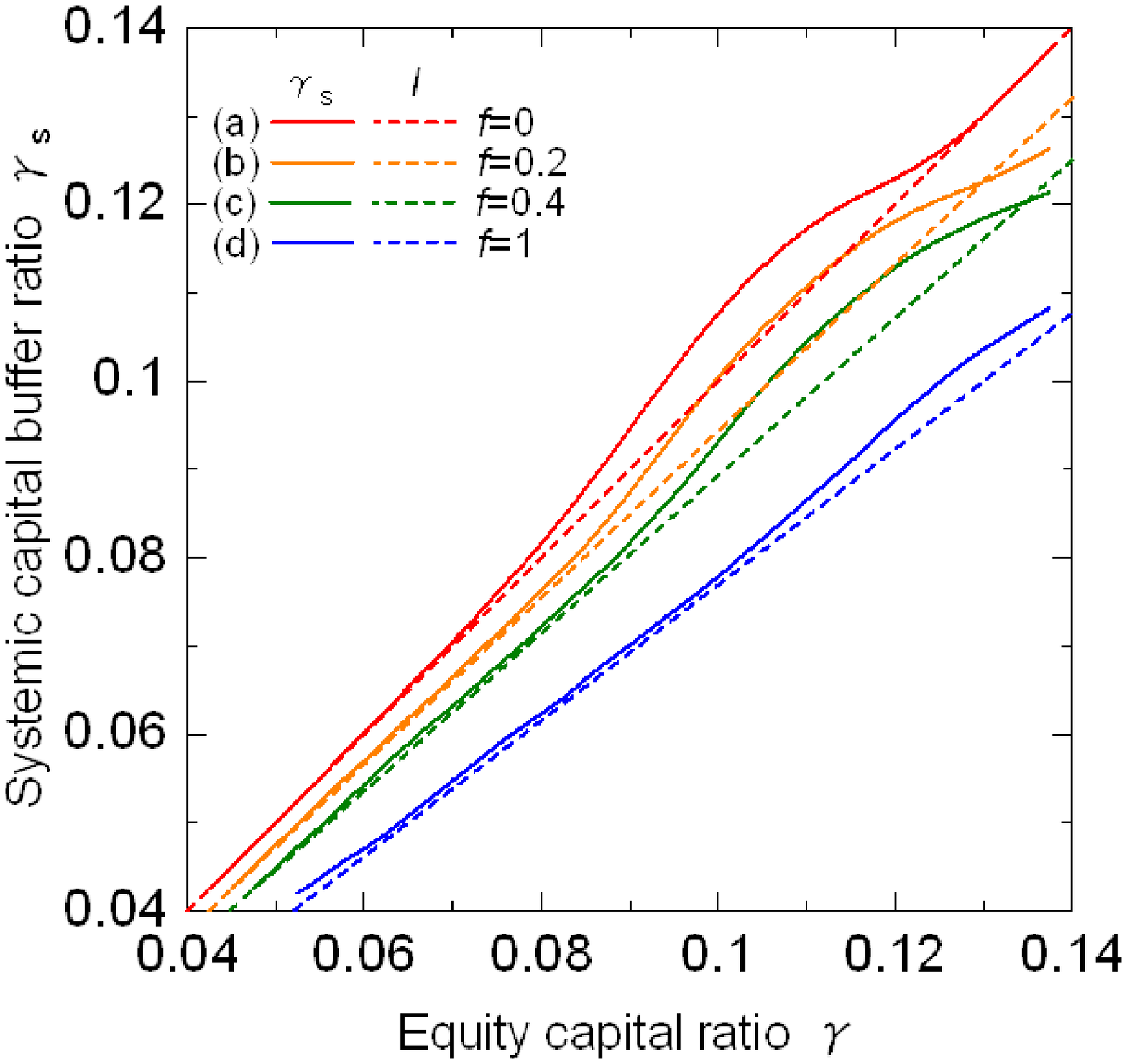}
\caption{Systemic capital buffer ratio $\gamma_{s}$ as a function of $\gamma$ for $\kappa=0.1$, $\rho=0.3$ (a more strongly connected interbank network), and $f=0$, 0.2, 0.4, and 1 when $N=500$, $M=2$, $S=10$, and $\theta=0.3$. The leverage ratio $l'$ are shown by broken lines.}
\vspace{0.2in}
\label{201309306}
\end{figure}

\begin{figure}
\centering
\includegraphics[width=3in, angle=-90]{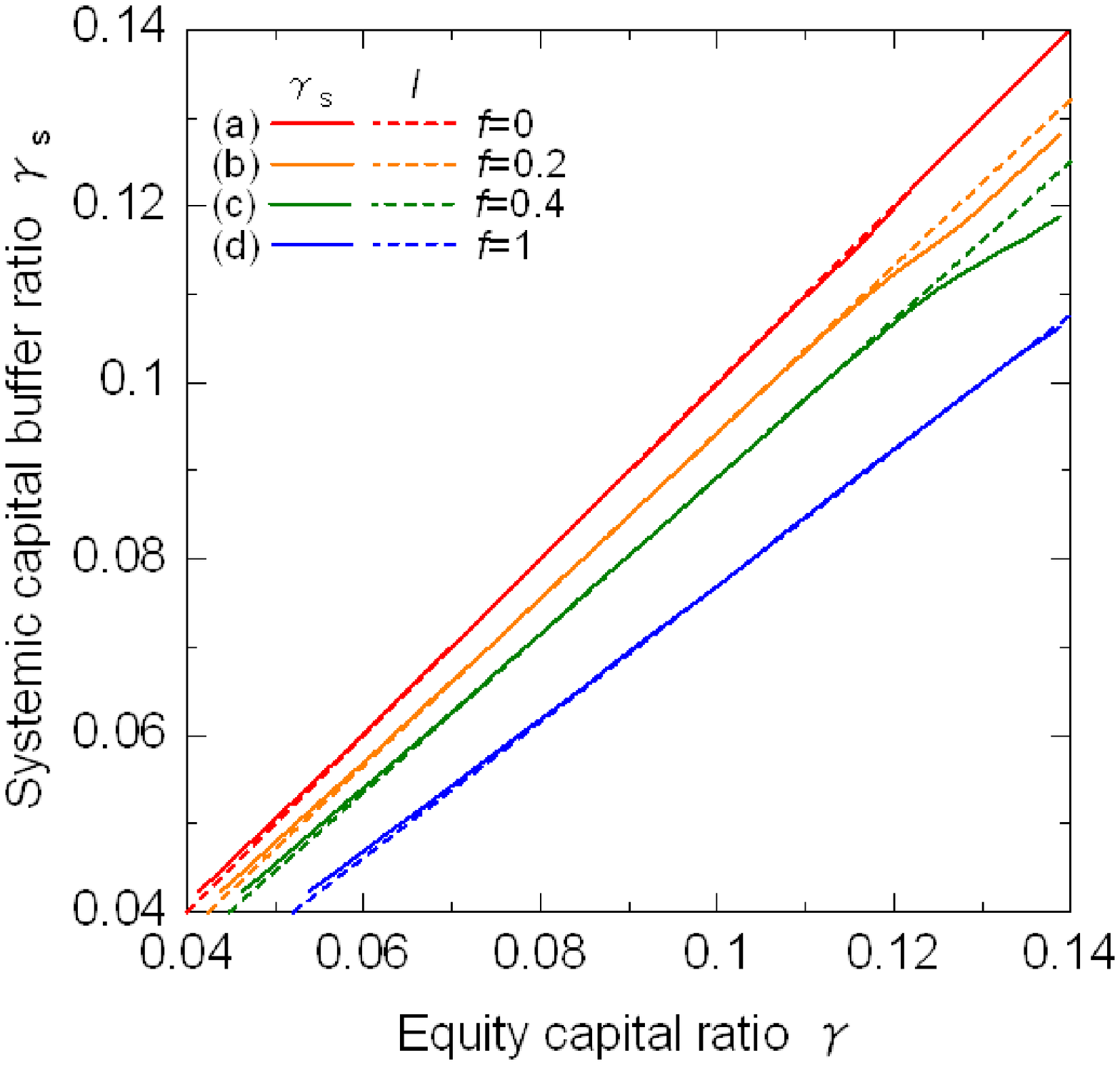}
\caption{Systemic capital buffer ratio $\gamma_{s}$ as a function of $\gamma$ for $\kappa=0.05$, $\rho=0.1$ (a less heavily concentrated interbank network), and $f=0$, 0.2, 0.4, and 1 when $N=500$, $M=2$, $S=10$, and $\theta=0.3$. The leverage ratio $l'$ are shown by broken lines.}
\label{201309307}
\end{figure}

\begin{figure}
\centering
\includegraphics[width=3in, angle=-90]{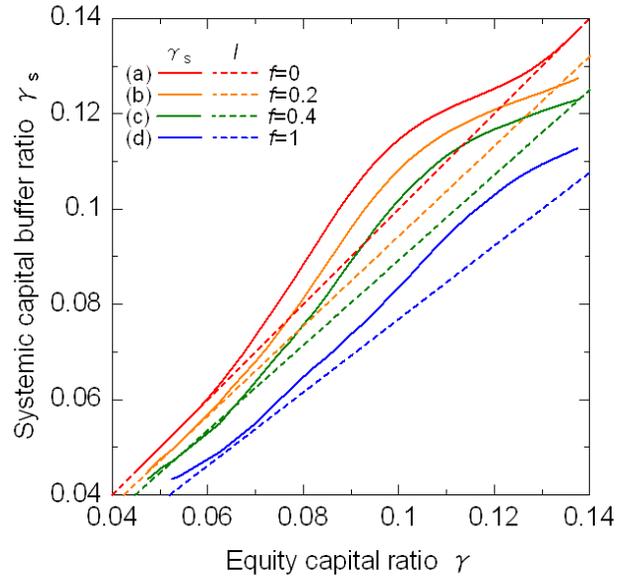}
\caption{Systemic capital buffer ratio $\gamma_{s}$ as a function of $\gamma$ for $\kappa=0.05$, $\rho=0.5$ (a more heavily concentrated interbank network), and $f=0$, 0.2, 0.4, and 1 when $N=500$, $M=2$, $S=10$, and $\theta=0.3$. The leverage ratio $l'$ are shown by broken lines.}
\label{201309308}
\end{figure}

The number of bank bankruptcies is measured under the experimental condition in \ref{Experimental} and the systemic capital buffer ratio is calculated in this section. Fig. \ref{201309303} shows the number of bank bankruptcies $F$ as a function of $\gamma$ as a fraction of $N$ when $\kappa=0.05$ and $\rho=0.3$. $F$ decreases gradually as $\gamma$ increases. Bank bankruptcies disappear at $\gamma=0.14$ in the financial system without risk transfer. $F$ in the financial system with risk transfer when $f=0$ (curve (a)) is nearly the same as that without risk transfer (curve (e)). This implies that credit default swaps have little impact on the severity of financial contagion. $F$ increases strikingly as $f$ increases. Additional loans impair the robustness of a financial system.

Fig. \ref{201309303} shows the systemic capital buffer ratio $\gamma_{s}$ under the same experimental condition as that for Fig. \ref{201309303}. The systemic capital buffer ratio is close to the leverage ratio $l'$ in eq.(\ref{l'}). The core tier 1 ratio $t'$ in eq.(\ref{t'}) is an optimistic estimate as an effective loss absorbency capability obviously since $t'=1.43l'$ for $f=0$ and $t'=1.3l'$ for $f=1$. Large leverage ratio is a vital backstop of a financial system while the core tier 1 ratio cannot be a predictor of the absolute severity of financial contagion. This implies that the risk transfer cannot make the risk weight of any interbank loans negligibly small.

Fig. \ref{201309305} shows the systemic capital buffer ratio $\gamma_{s}$ for $\kappa=0.01$ and $\rho=0.3$ (a less strongly connected interbank network). Fig. \ref{201309306} shows the systemic capital buffer ratio $\gamma_{s}$ when $\kappa=0.1$ and $\rho=0.3$ (a more strongly connected interbank network). The denseness of an interbank network does not affect the impact of the risk transfer. The leverage ratio is still a good estimate as an effective loss absorbency capability. 

Fig. \ref{201309307} shows the systemic capital buffer ratio $\gamma_{s}$ when $\kappa=0.05$ and $\rho=0.1$ (a less heavily concentrated interbank network). The top 1 percent banks own 10 percent of the total interbank loans, and equivalently, about 10 percent of the total assets. The curves in this figure are very close to those in Fig. \ref{201309305}. Fig. \ref{201309306} shows the systemic capital buffer ratio $\gamma_{s}$ when $\kappa=0.05$ and $\rho=0.5$ (a more heavily concentrated interbank network). The top 1 percent banks own 50 percent of the total interbank loans. When the network is heavily concentrated, the systemic capital buffer ratio is bigger than the leverage ratio around $\gamma=0.1$. The risk transfer from small and medium banks to big banks mitigates the severity of financial contagion. Although the CDS works under this condition, the core tier 1 ratio is still an optimistic estimate as an effective loss absorbency capability.

\section{Discussion}

The key findings of this study are as follows. The leverage ratio is a good estimate of a systemic capital buffer ratio as the backstop of a financial system. The risk transfer from small and medium banks to big banks in an interbank network does not mitigate the severity of financial contagion except for a heavily concentrated interbank network. Additional interbank loans after the risk transfer undermine the robustness of a financial system. Analysts have criticized that banks do not raise fresh equity capital but just optimize risk weighted assets to comply with the core tier 1 ratio requirement. They remain highly leveraged. The focus of the international Basel III capital requirements has shifted from the core tier 1 ratio to the leverage ratio recently. In July 2013, Federal Reserve unveiled the implementation of the international Basel III capital requirements which request banks in US to hold a higher level of capital against the overall assets. The judgment of controversial risk weights is stripped out by the leverage ratio. This study demonstrates that the shift of the focus is reasonable from the view point of a systemic capital buffer ratio.

The experimental conditions in \ref{Experimental} are a typical example. But experimental conditions are not limited to those. The interbank network topology $\mbox{\boldmath{$Z$}}$ may be a core-periphery model\cite{Craig} or other highly heterogeneous models. The risk transfer pattern $\mbox{\boldmath{$Y$}}$ and the investment portfolio $\mbox{\boldmath{$X$}}$ may be non-random. There may be a number of bank categories which have different strategies in the investment and risk transfer. The prices of the external asset classes $\mbox{\boldmath{$v$}}$ may obey a logarithmic normal distribution or other long tailed multi-variate distributions. Investigating these experimental conditions are for future works. Further extension of the asset network systemic risk model (ANWSER) is necessary to study more practical financial systems including (1) such detailed terms as maturity of interbank loans, CDS contracts, and similar financial derivatives\cite{Karim}, (2) the role of clearing houses and netting of exposures, (3) withdrawal of loans, refinancing, and other means to raise money\cite{Duffie}, (4) liquidity of assets, correlation between the prices of external asset classes, and other market mechanisms. Other interesting research topics are the bank run in a fractional reserve banking when a financial crisis precipitates a banking panic and the realistically dynamic evolution of the interbank network.

Another issue is that the interbank network topology, risk transfer pattern, and investment portfolio are not static in reality and cannot be observed directly either. In the study of epidemic contagion in social network analysis, the transportation network and relevant parameters are inferred from the observation on the number of patients with a statistical analysis\cite{Maeno11}, \cite{Maeno10}. An impending epidemic outbreak is to be predicted. Similarly, it may be possible to foresee financial contagion by inferring the basic structure of the interbank network topology, risk transfer pattern, and investment portfolio from statistical datasets and real time monitoring. Such an analysis aids the supervisors and other relevant authorities in designing the financial system theoretically and in making regulatory policies in practice. This is the goal of an emerging field of systems economics.

\section*{Acknowledgment}
The authors would like to thank Hidetoshi Tanimura, Ernst\&Young ShinNihon LLC, and Akira Namatame, National Defense Academy, for their advice and discussion.

\end{document}